\documentclass[aps,pre,reprint,showpacs,superscriptaddress,nofootinbib,groupaddress,showkeys,floatfix,byrevtex]{revtex4-1}
\usepackage{amsmath}
\usepackage{amsfonts}
\usepackage{amssymb}
\usepackage[dvipsnames]{xcolor}
\usepackage{graphicx,color,subeqnarray}
\usepackage{dcolumn}
\usepackage{bm}
\usepackage{srcltx}
\usepackage{soul}
\usepackage{appendix}
\begin{document}

\title{Anomalous diffusion of self-propelled particles}

\author{Francisco J. Sevilla}
\email{fjsevilla@fisica.unam.mx}
\thanks{corresponding author}
\affiliation{Instituto de F\'isica, Universidad Nacional Aut\'onoma de M\'exico,
Apdo.\ Postal 20-364, 01000, Ciudad de M\'exico, M\'exico
}

\author{Guillermo Chacón-Acosta}
\email{gchacon@cua.uam.mx}
\affiliation{Departamento de Matemáticas Aplicadas y Sistemas, Universidad Autónoma Metropolitana-Cuajimalpa, Vasco de Quiroga 4871, Santa Fe, Cuajimalpa, Mexico City, 05348, Mexico}
\affiliation{Institute of Physics \& Astronomy, University of Potsdam, D-14776 Potsdam-Golm, Germany}

\author{Trifce Sandev}
\email{trifce.sandev@manu.edu.mk}
\affiliation{Research Center for Computer Science and Information Technologies,
Macedonian Academy of Sciences and Arts, Bul. Krste Misirkov 2, 1000 Skopje, Macedonia}
\affiliation{Institute of Physics, Faculty of Natural Sciences and Mathematics,
Ss.~Cyril and Methodius University, Arhimedova 3, 1000 Skopje, Macedonia}
\affiliation{Institute of Physics \& Astronomy, University of Potsdam, D-14776 Potsdam-Golm, Germany}

\date{\today}

\begin{abstract}
The transport equation of active motion is generalised to consider time-fractional dynamics for describing the anomalous diffusion of self-propelled particles observed in many different systems. In the present study, we consider an arbitrary active motion pattern modelled by a scattering function that defines the dynamics of the change of the self-propulsion direction. The exact probability density of the particle positions at a given time is obtained. From it, the time dependence of the moments, i.e., the mean square displacement and the kurtosis for an arbitrary scattering function, are derived and analysed. Anomalous diffusion is found with a crossover of the scaling exponent from $2\alpha$ in the short-time regime to $\alpha$ in the long-time one, $0<\alpha<1$ being the order of the fractional derivative considered. It is shown that the exact solution found satisfies a fractional diffusion equation that accounts for the non-local and retarded effects of the Laplacian of the probability density function through a coupled temporal and spatial memory function. Such a memory function holds the complete information of the active motion pattern. In the long-time regime, space and time are decoupled in the memory function, and the time fractional telegrapher's equation is recovered. Our results are widely applicable in systems ranging from biological microorganisms to artificially designed self-propelled micrometer particles.
\end{abstract}

\maketitle
\section{Introduction}
The anomalous transport of either passive or active particles in crowded environments \cite{HoflingRepProgPhys2013,RevereySciReps2015,BechingerRMP2016,ArgunPRE2016,BasakBioPhysRev2019} has been observed in the form of anomalous diffusion, which describes the non-linear grow in time of the mean squared displacement (MSD), i.e., $\langle\boldsymbol{x}^2(t\rangle)$ scales with time as the power law $t^{\alpha}$ with $\alpha>0$~\cite{metzler2000random,*metzler2004restaurant}. Of particular interest is the crossover from power-law time scaling in the short-time regime, to sub-diffusion in the long-time one. This crossover has been predicted in computer simulations of the motion of the centre of mass of a polymer chain diffusing in an active bath \cite{VandebroekPRE2015}; in the diffusion of an active particle in polymer solutions \cite{YuanPhysChemChemPhys2019}; and in
the diffusive behaviour of a tracer particle analysed in numerical simulations in a lattice model of a crowded active
environment \cite{AbbaspourPRE2021}. 

Active motion is characterised by being \emph{persistent}, and deep consequences due to persistence have emerged in the form of diverse nonequilibrium collective dynamics: the well-known motility induced phase separation, and more recently, the anomalous segregation dynamics \cite{MonesNewJPhys2015,MoranNewJPhys2022}; even at the single particle level active motion has been of great interest \cite{SquarciniNewJPhys2022}. On the other hand, besides persistence, different intrinsic nonequilibrium features have been added to the motion of self-propelled particles \cite{LemaitreNewJPhys2023}.
Different models of active motion depends on the specific pattern of self-motility. Notwithstanding this, the so-called \emph{run-and-tumble} and \emph{active Brownian} motion have been widely used to analyse transport properties of active matter. The stochastic dynamics of run-and-tumble particles simply alternates stochastic time periods during which the particle moves along a randomly chosen direction. If the distribution of such time periods is exponential and characterised by time scale $\Lambda^{-1}$, we have that probability density of the particle being located at position $\boldsymbol{x}$, moving along the direction $\hat{\boldsymbol{v}}=(\cos\varphi,\sin\varphi)$ at time $t$ satisfies the equation~\cite{SchnitzerPRE1993,PorraPRE1997,SevillaPRE2020}
\begin{multline}\label{RT-Equation}
\frac{\partial}{\partial t}\mathcal{P}({\boldsymbol{x}},\varphi ,t)+v_{0}\hat{\boldsymbol{v}}\cdot \nabla \mathcal{P}({\boldsymbol{x}},\varphi
,t)\\
=\Lambda\int_{-\pi}^{\pi}d\varphi^{\prime}Q(\varphi-\varphi^{\prime})\mathcal{P}(\boldsymbol{x},\varphi^{\prime},t)-\Lambda\mathcal{P}(\boldsymbol{x},\varphi,t),
\end{multline}
where $Q(\varphi)$ denotes the probability density for making a change  by an angle $\varphi$ on the direction of motion and carries out the specifics of the pattern of active motion (run-and-tumble \cite{SevillaPRE2020}). $\Lambda^{-1}$ gives the average time the particle moves in a particular direction $\varphi$. An analysis from this equation shows a crossover from ballistic diffusion in the short-time regime, to normal diffusion in the long-time regime independent of the particular pattern of active motion embedded in $Q(\varphi)$,~\cite{SevillaPRE2020}.

In order to describe the presence of anomalous diffusion of particles that move persistently with an arbitrary pattern of active motion, in this paper we generalise Eq.~\eqref{RT-Equation} by promoting the first-order time derivative to its fractional counterpart. This theoretical framework models the effects of crowded environments on the diffusion, usually anomalous, of active particles. 

The paper is organised as follows. In Section II we consider a generalisation of Eq. \eqref{RT-Equation}, in which the retarded effects of the rate of change of the probability density are taken into consideration by a power-law memory function. Our main results focus on the probability density function of an active particle located at position $\boldsymbol{x}$ at time $t$ (independent of the direction of motion) for which we give an exact analytical expression in Fourier-Laplace variables. 
As a special cases we recover the time fractional diffusion, wave and telegrapher's equations. A detailed study of the moments of the fundamental solution of the equation are given in Section III. We calculate the second and fourth moments and analysed the kurtosis to characterised the transport properties of active motion under fractional dynamics. A crossover in the time-scaling exponent that characterises anomalous diffusion is observed form our analysis of the MSD. The special case of $Q(\varphi)=1/2\pi$, the uniform angle-scattering distribution, is considered. Finally, in Section IV we give summary of the obtained results and some remarks for further investigation. 

\section{\label{SectModel}The two-dimensional active transport equation}
If the effects of the nonequilibrium crowded environment anomalously slows down the diffusion of the active particle we make the replacement
\begin{equation}\label{fracD}
\frac{\partial}{\partial t}\longrightarrow\tau^{\alpha-1}\frac{\partial^{\alpha}}{\partial t^{\alpha}}
\end{equation}
where 
\begin{equation}\label{caputo derivative}
\frac{\partial^{\alpha}}{\partial t^{\alpha}}\equiv\frac{1}{\Gamma(1-\alpha)}\int_{0}^{t}ds \frac{1}{(t-s)^{\alpha}}\frac{\partial}{\partial s},
\end{equation}
is the Caputo fractional derivative of order $0<\alpha\le1$~\cite{prudnikov2003integrals}, and $\tau$ is an arbitrary time scale that gives the proper physical units when $\partial/\partial t$ is replaced by $\partial^\alpha /\partial t^\alpha$ and that will be fixed.

Such time fractional derivative have been used to model anomalous diffusive processes of passive particles, for which the mean squared displacement scales with time as the power law time~\cite{metzler2000random}
\begin{equation}\label{anomalous diffusion msd}
    \langle x^2(t)\rangle\sim t^{\alpha}.
\end{equation}
If $0<\alpha<1$ the corresponding process is subdiffusive, while for $\alpha>1$ -- superdiffusive. For $\alpha=1$ one recovers the normal diffusive process. The time fractional derivatives naturally occur in different stochastic processes. For example, if in the continuous time random walk model one considers a long-tailed waiting time ($\psi(t)\sim t^{-\alpha-1}$, $0<\alpha<1$) of the particle the governing Fokker-Planck equation is with time fractional derivative of order $\alpha$~(\ref{caputo derivative}), see Ref.~\cite{metzler2000random}. The resulting process is subdiffusive with MSD of form~(\ref{anomalous diffusion msd}) since $0<\alpha<1$. The physical description of a diffusion process in a constrained geometries, such as comb structures, also leads to fractional Fokker-Planck equation with a time fractional derivative~(\ref{caputo derivative}) of order $1/2$ as a governing equation for the particle movement along the main channel, see Refs.~\cite{iomin2018fractional,sandev2023special}. Another example is the fractional Langevin equation~\cite{lutz2001fractional} with time fractional derivative in the friction term for a particle moving in a viscous medium. The medium is characterised by a friction memory kernel, and the particle is driven by a fractional Gaussian noise (a noise with Gaussian amplitude and power-law correlations) which correlation is related to the friction memory kernel via the second fluctuation-dissipation theorem~\cite{kubo}. Moreover, it is shown that such fractional friction can be used to explain the cage effects as a viscoelastic property of the medium~\cite{burov2008critical,burov2008fractional}, where the rattling motion of the surrounding particles are captured by the power-law friction memory kernel. We also mention the time fractional telegrapher's equation with time fractional derivative which describes a persistent random walk, a two-state random walk, with long-tailed waiting time~\cite{masoliver2016fractional}. All these processes show anomalous diffusive behaviour. 

By use of the prescription~\eqref{fracD}, the equation that gives the change in time for the probability density, $\mathcal{P}({\boldsymbol{x}},\varphi ,t)$, of a single particle being at position $\boldsymbol{x}$, moving at constant speed $v_{0}$ along a direction given by the angle $\varphi$ at time $t$ is now
\begin{multline}\label{FRT-Equation}
\tau^{\alpha-1}\frac{\partial^{\alpha} }{\partial t^{\alpha}}\mathcal{P}({\boldsymbol{x}},\varphi ,t)+v_{0}\hat{\boldsymbol{v}}\cdot \nabla \mathcal{P}({\boldsymbol{x}},\varphi
,t)\\
=\Lambda\int_{-\pi}^{\pi}d\varphi^{\prime}Q(\varphi-\varphi^{\prime})\mathcal{P}(\boldsymbol{x},\varphi^{\prime},t)-\Lambda\mathcal{P}(\boldsymbol{x},\varphi,t),
\end{multline}
the unit vector $\hat{\boldsymbol{v}}$ is defined by $(\cos\varphi,\sin\varphi)$, $\varphi$ being the angle between the direction of motion and the horizontal axis of a given Cartesian reference frame.

We are interested in analysing $\mathcal{P}(\boldsymbol{x},\varphi,t)$ when the initial condition $\mathcal{P}({\boldsymbol{x}},\varphi ,0)=\delta ^{(2)}({\boldsymbol{x}})/2\pi $ is considered. This corresponds to the case of an ensemble of independent active particles that depart from the origin in a Cartesian system of coordinates, and propagates in a random direction of motion drawn from the uniform distribution in $[-\pi,\pi]$, $\delta ^{(2)}({\boldsymbol{x}})$ being the two dimensional Dirac's delta function. 

Spatial isotropy of the system allows to apply the Fourier transform to Eq.~(\ref{FRT-Equation}) and obtain 
\begin{multline}\label{Transformed}
\tau^{\alpha-1}\frac{\partial^{\alpha} }{\partial t^{\alpha}}\widetilde{\mathcal{P}}({\boldsymbol{k}},\varphi ,t)+iv_{0}\,\hat{\boldsymbol{v}}\cdot {\boldsymbol{k}}\, \widetilde{\mathcal{P}}({\boldsymbol{k}},\varphi ,t) \\
=\Lambda\int_{-\pi}^{\pi}d\varphi^{\prime} Q\left(\varphi-\varphi^{\prime}\right)\widetilde{\mathcal{P}}(\boldsymbol{k},\varphi^{\prime},t)\\
-\Lambda \widetilde{\mathcal{P}}(\boldsymbol{k},\varphi,t),
\end{multline}
where
\begin{equation}
\widetilde{\mathcal{P}}({\boldsymbol{k}},\varphi ,t)=\int \frac{d^{2}x}{2\pi}\,e^{-i\boldsymbol{k}\cdot \boldsymbol{x}}\,\mathcal{P}(\boldsymbol{x},\varphi ,t),
\label{fourier}
\end{equation}
denotes the symmetric Fourier transform of $\mathcal{P}(\boldsymbol{x},\varphi,t)$ and ${\boldsymbol{k}}=(k_{x},k_{y})$, denotes the system's wave-vector. The following Fourier 
series expansion,
\begin{equation}\label{Pexpansion}
\widetilde{\mathcal{P}}({\boldsymbol{k}},\varphi ,t)=\frac{1}{2\pi}\sum\limits_{n=-\infty}
^{\infty }\widetilde{p}_{n}({\boldsymbol{k}},t)\, e^{in\varphi},  
\end{equation}
is suitable since it fulfils the periodicity condition of the probability density, $\widetilde{\mathcal{P}}({\boldsymbol{k}},\varphi 
,t)=\widetilde{\mathcal{P}}({\boldsymbol{k}},\varphi+2\pi,t)$.

The coefficients $\widetilde{p}_{n}(\boldsymbol{k},t)$ in the expansion~\eqref{Pexpansion} are obtained by the use of the 
standard orthogonality relation among the Fourier basis functions $\left\{e^{in\varphi}\right\}$, explicitly
\begin{align}
\widetilde{p}_{n}({\boldsymbol{k}},t)&=\int \frac{d^{2}x}{2\pi}\, e^{-i\boldsymbol{k}\cdot\boldsymbol{x}}\, p_{n}(\boldsymbol{x},t)\nonumber\\
&=\int_{-\pi}^{\pi}d\varphi
\, \widetilde{\mathcal{P}}({\boldsymbol{k}},\varphi ,t)e^{-in\varphi}.  \label{inv}
\end{align}

After substitution of Eq.~\eqref{Pexpansion} into Eq.~\eqref{Transformed}, and use of the orthogonality of the Fourier basis functions, a set of coupled 
ordinary differential equations for the coefficients $\widetilde{p}_{n}({
\boldsymbol{k}},t)$ is obtained, namely
\begin{multline}\label{Hierarchy}
\tau^{\alpha-1}\frac{d^{\alpha}}{dt^{\alpha}}\widetilde{p}_{n}(\boldsymbol{k},t) + \Lambda\lambda_{n}\widetilde{p}_{n}(\boldsymbol{k},t)\\
=-\frac{v_{0}}{2}ik\bigl[
e^{-i\theta}\, \widetilde{p}_{n-1}(\boldsymbol{k},t)
+e^{i\theta}\, \widetilde{p}_{n+1}(\boldsymbol{k},t)\bigr],
\end{multline}
where $\theta$ and $k$ correspond to the polar coordinates of the two-dimensional Fourier vector $\boldsymbol{k}$, i.e.,  $k_{x}\pm ik_{y}=ke^{\pm i\theta }$. Equations~\eqref{Hierarchy} are complemented by the initial conditions $\widetilde{p}_{n}^{(0)}(\boldsymbol{k})=(2\pi)^{-1}\delta_{n,0}$, which are obtained straightforwardly 
from the initial distribution considered, i.e., $\mathcal{P}(\boldsymbol{x},\varphi,0)=\delta^{(2)}(\boldsymbol{x})/2\pi$. $\lambda_{n}$ is a complex number given by
\begin{equation}
 \lambda_{n}=1-\langle e^{-in\varphi}\rangle_{Q},
\end{equation}
where
\begin{equation}
 \langle e^{-in\varphi}\rangle_{Q}\equiv\int_{-\pi}^{\pi}d\varphi\, Q(\varphi)e^{-in\varphi}.
\end{equation}

Accordingly, the main features of a particular pattern of active motion are encoded in the distribution of scattered angles  $Q(\varphi)$, which 
entails the particular orientation dynamics of the swimming direction. Such features are equivalently inherited in the trigonometric moments:
\begin{subequations}\label{lambda-n}
 \begin{align}
  \Gamma_{n}&=1-\langle\cos n\varphi\rangle_{Q},\label{Gamma-n}\\
  \Omega_{n}&=\langle\sin n\varphi\rangle_{Q},\label{Omega-n}
  \end{align}
\end{subequations}
which correspond to the real and imaginary part of $\lambda_{n}$, respectively, thus $\lambda_{n}=\Gamma_{n}+i\, \Omega_{n}$. These quantities fully characterise the statistical properties of active motion.

The following properties for $\Gamma_{n}$ and $\Omega_{n}$ can be deduced in a straightforward way. From the 
normalisation of $Q$ we have that $\Gamma_{0}=\Omega_{0}=0$, and since $Q(\varphi)$ is a real-valued function, we have that the complex conjugate of 
$\lambda_{n}$ is given by $\lambda_{n}^{*}=\lambda_{-n}$, which implies $\Gamma_{n}=\Gamma_{-n}$ and $\Omega_{n}=-\Omega_{-n}$. Notice further that $0\le\Gamma_{n}\le2$ 
and that $-1\le\Omega_{n}\le1$. 

Laplace transform of Eq.~\eqref{Hierarchy} gives
\begin{multline}\label{HierarchyII}
\Bigl[\tau^{\alpha}\epsilon^{\alpha}+ \tau\Lambda\lambda_{n}\Bigr]\widetilde{p}_{n}(\boldsymbol{k},\epsilon)=\tau^{\alpha}R_{n,\alpha}(\boldsymbol{k}, \epsilon)\\
 -\tau\frac{v_{0}}{2} ik\Bigl[e^{-i\theta}\widetilde{p}_{n-1}(\boldsymbol{k},\epsilon)+e^{i\theta}\widetilde{p}_{n+1}(\boldsymbol{k},\epsilon)\Bigr].
\end{multline}
The unnecessary additional information given by the arbitrary ratio of time scales $\tau/\Lambda^{-1}$ is reduced by fixing such ratio to unity, thus we fix $\tau=\Lambda^{-1}$ in this work. In~\eqref{HierarchyII} we have $R_{n,\alpha}(\boldsymbol{k},\epsilon)$ encompasses the initial data required by the order of the fractional derivative, $\alpha$, namely~\cite{PodlubnyBook}
\begin{equation}
 R_{n,\alpha}(\boldsymbol{k},\epsilon)=\epsilon^{\alpha-1}\bigl[\widetilde{p}_{n}(\boldsymbol{k},t)\bigr]_{t=0},
\end{equation}
which for the initial configuration considered, we have $R_{n,\alpha}(\boldsymbol{k},\epsilon)=\epsilon^{\alpha-1}\delta_{n,0}$, with $0<\alpha<1$. By recursively solving Eqs.~\eqref{HierarchyII} (see Ref.~\cite{SevillaPRE2020}), we get an exact expression for $\widetilde{p}_0(\boldsymbol{k},\epsilon)$ in terms of continued fractions
\begin{widetext}
\begin{equation}\label{p0}
\widetilde{p}_{0}(\boldsymbol{k},\epsilon)=\cfrac{(\epsilon^{\alpha-1}/\Lambda^{\alpha})}{(\epsilon/\Lambda)^{\alpha}
	  + \cfrac{(v_{0}/2\Lambda)^{2}\boldsymbol{k}^{2}}{(\epsilon/\Lambda)^{\alpha}+\lambda_{1}
          + \cfrac{(v_{0}/2\Lambda)^{2}\boldsymbol{k}^{2}}{(\epsilon/\Lambda)^{\alpha}+\lambda_{2}
          + \cfrac{(v_{0}/2\Lambda)^{2}\boldsymbol{k}^{2}}{(\epsilon/\Lambda)^{\alpha}+\lambda_3+\ddots
           } } }+
	    \cfrac{(v_{0}/2\Lambda)^{2}\boldsymbol{k}^{2}}{(\epsilon/\Lambda)^{\alpha}+\lambda_{1}^{*}
          + \cfrac{(v_{0}/2\Lambda)^{2}\boldsymbol{k}^{2}}{(\epsilon/\Lambda)^{\alpha}+\lambda_{2}^{*}
          + \cfrac{(v_{0}/2\Lambda)^{2}\boldsymbol{k}^{2}}{(\epsilon/\Lambda)^{\alpha}+\lambda_{3}^{*}+\ddots
           } }}
           },
\end{equation}
for which the role of the pattern of active motion is marked by the dependence on $\lambda_n, \lambda_n^*$, $n=1,2,\ldots$, and on $v_{0}$.  

\subsection{The generalised time fractional diffusion equation.}
Last solution can be cast in the form
\begin{equation}\label{Solution-D}
\widetilde{p}_0(\boldsymbol{k},\epsilon)=\frac{\epsilon^{\alpha-1}}{\epsilon^{\alpha}+\Lambda^\alpha(v_{0}/2\Lambda)^{2}\, \widetilde{\mathfrak{D}}_\alpha(\boldsymbol{k},\epsilon)\,  \boldsymbol{k}^2}
\end{equation}
(see Ref.~\cite{SevillaPRE2020,SevillaIJMPB2022} for the case $\alpha=1$),
which after rearrangement it can be written in spatial and time variables as the fractional diffusion equation
\begin{equation}\label{Solution-D-InvLF}
\frac{\partial^{\alpha}}{\partial t^{\alpha}}p_0(\boldsymbol{x},t)=\Lambda^\alpha\biggl(\frac{v_0}{2\Lambda}\biggr)^2\int_{0}^{t} ds\int d^{2}x^{\prime}\, \mathfrak{D}_{\alpha}(\boldsymbol{x}-\boldsymbol{x}^{\prime},t-s){\nabla^{\prime}}^{2}p_0(\boldsymbol{x}^{\prime},s).
\end{equation}
This takes into account the non local and retarded effects of the Laplacian of $p_0(\boldsymbol{x,t})$ into the fractional dynamics through the memory function, coupled in time and space, $\mathfrak{D}_\alpha(\boldsymbol{x},t)$. The connecting function $\mathfrak{D}(\boldsymbol{x},t)$ is obtained as the inverse Fourier-Laplace transform of 
\begin{equation}\label{ConnectingF}
\widetilde{\mathfrak{D}}_\alpha(\boldsymbol{k},\epsilon)=
	  \cfrac{1}{(\epsilon/\Lambda)^{\alpha}+\lambda_{1}
          + \cfrac{(v_{0}/2\Lambda)^{2}\boldsymbol{k}^{2}}{(\epsilon/\Lambda)^{\alpha}+\lambda_{2}
          + \cfrac{(v_{0}/2\Lambda)^{2}\boldsymbol{k}^{2}}{(\epsilon/\Lambda)^{\alpha}+\lambda_{3}+\ddots
           } } }+
	    \cfrac{1}{(\epsilon/\Lambda)^{\alpha}+\lambda_{1}^{*}
          + \cfrac{(v_{0}/2\Lambda)^{2}\boldsymbol{k}^{2}}{(\epsilon/\Lambda)^{\alpha}+\lambda_{2}^{*}
          + \cfrac{(v_{0}/2\Lambda)^{2}\boldsymbol{k}^{2}}{(\epsilon/\Lambda)^{\alpha}+\lambda_{3}^{*}+\ddots
           } }}.
\end{equation}
\end{widetext}

The active motion described here, and whose synthesis is encapsulated in Eq.~\eqref{Solution-D}, reduces in the long-time regime to the well-known time fractional diffusion equation (fTDE) (\cite{SchneiderJMathPhys1989,MainardiChaosSolFrac1996,MetzlerPhysReps2000} and to the time fractional telegrapher's equation (fTTE)~\cite{masoliver2016fractional} as is shown in the following paragraphs. The fTDE is directly obtained from \eqref{Solution-D} after taking the asymptotic limit for which $\widetilde{\mathfrak{D}}(\boldsymbol{k},\epsilon)\xrightarrow[]{}2$. Thus Eq. \eqref{Solution-D} reduces, after inversion of the Fourier-Laplace transform to
\begin{equation}\label{fTDE}
\frac{\partial^\alpha}{\partial t^\alpha}p_\textrm{fTDE}(\boldsymbol{x},t)=\Lambda^\alpha\frac{v_0^2}{2\Lambda^2}\nabla^2p_\textrm{fTDE}(\boldsymbol{x},t).
\end{equation}

As occurs in the case $\alpha=1,$ in Ref. ~\cite{SevillaPRE2020}, the truncation of $\widetilde{\mathfrak{D}}(\boldsymbol{k},\epsilon)$ to the first approximant (some times called the P$_1$ approximation of the transport equation \cite{HeizlerNucSciEng2010,Espinos-ParedesNucSciEng2012}), i.e., 
\begin{equation}\label{1stApproximantLaplace}
\widetilde{\mathfrak{D}}_\textrm{fTTE}(\epsilon)\equiv\widetilde{\mathfrak{D}}(\boldsymbol{k},\epsilon)\bigl\vert_{\boldsymbol{k}=0}
=\frac{\Lambda^{\alpha}}{\epsilon^\alpha+\Lambda^{\alpha}{\lambda}_1}+\frac{\Lambda^{\alpha}}{\epsilon^{\alpha}+\Lambda^{\alpha}{\lambda}^*_1}
\end{equation}
is $\boldsymbol{k}$ independent and therefore local in space. This leads to the generalised time fractional telegrapher's equation 
\begin{multline}\label{GralfracTimeTE}
\frac{\partial^\alpha}{\partial t^\alpha}p_\textrm{fTTE}(\boldsymbol{x},t)=\Lambda^{\alpha}\bigg(\frac{v_0^2}{2\Lambda}\biggr)^2\times\\
\int_{0}^{t}ds\, \mathfrak{D}_\textrm{fTTE}(t-s)
\nabla^{2}p_\textrm{fTTE}(\boldsymbol{x},s),
\end{multline}
where 
\begin{equation}\label{1stApproximantTime}
\mathfrak{D}_\textrm{fTTE}(t)=\Lambda^\alpha t^{\alpha-1}
\bigl[E_{\alpha,\alpha}\bigl(-\lambda_1\Lambda^\alpha t^\alpha\bigr)+E_{\alpha,\alpha}\bigl(-{\lambda_1}^*\Lambda^\alpha t^\alpha\bigr)\Bigr],
\end{equation} 
and $E_{\rho,\beta}(z)$ is the two-parameters Mittag-Leffler function, see Appendix \ref{appA}.
In the case $\alpha=1$, the memory function $\mathfrak{D}_\textrm{fTTE}(\boldsymbol{x},t)$ can be factorised into the factors $2e^{-\Lambda\Gamma_1 t}\cos(\Lambda\Omega_1 t)$ \cite{SevillaPRE2020} [we have used that $E_{1,1}(-z)=e^{-z}$ and  $\lambda_1=\Gamma_1+i\, \Omega_1$], and reveals the meaning of the time scales $t_\textrm{pers}=(\Lambda\Gamma_1)^{-1}$ and $\omega_\textrm{pers}=\Lambda\Omega_1$, associated to two antagonistic effects embedded in $Q(\varphi)$. The first $t_\textrm{pers}$, defines the effective persistence time of the swimming direction and form this the persistence length $l_\textrm{pers}=v_0t_\textrm{pers}$. The second, $\omega_\textrm{pers}$, 
exhibits the rotational bias or effective angular velocity that takes into account a systematic change of the swimming direction. For $0<\alpha<1$ no such a factorisation is possible, thus $t_\textrm{pers}$ and $\omega_\textrm{pers}$ are intertwined, this is observed in the mean squared displacement as will be discussed in the following sections.  

For angle distributions $Q(\varphi)$ symmetric about $\varphi=0$, we have $\Omega_n=0$ for all $n$, therefore no rotational bias ($\Omega_1=0$) is present in the dynamics, and $\mathfrak{D}_\textrm{fTTE}(t)$ simplifies to $2t^{\alpha-1}E_{\alpha,\alpha}\bigl(-\Gamma_1\Lambda^\alpha t^\alpha\bigr)$, which leads to the two-dimensional time fractional telegrapher equation
\begin{multline}\label{fTTEq}
\frac{\partial^{2\alpha}}{\partial t^{2\alpha}}p_\textrm{fTTE}(\boldsymbol{x},t)+\Lambda^{\alpha}\Gamma_1\frac{\partial^\alpha}{\partial t^\alpha}p_\textrm{fTTE}(\boldsymbol{x},t)=\\
\Lambda^{2(\alpha-1)}\frac{v_0^2}{2}\nabla^{2}p_\textrm{fTTE}(\boldsymbol{x},t).
\end{multline}
The one-dimensional version of this equation has received vast attention during the the last three decades, since the seminal work of Compte and Metzler in Ref.~\cite{compte1997} as one of the possible generalisations to the Cattaneo equation, to more recent derivations as the ones presented in Refs.~\cite{masoliver2016fractional,GorzkaPhysRevE2020}. Notice that the fTTE derived here~\eqref{GralfracTimeTE} accounts for the effects of the rotational bias ($\Omega_1\neq 0$) from the scattering function $Q(\varphi)$; moreover, by considering more approximants to the connecting function \eqref{ConnectingF}, more time scales could also be accounted for.

In the short-time regime,  coherent motion derives from Eq.~\eqref{FRT-Equation} by considering that $\epsilon$ is larger than any of the $\Gamma_n$'s and $\Omega_n$'s that characterise the pattern of active motion, thus Eq. (\ref{p0}) reduces to
\begin{equation}\label{p0-ShortTR}
\widetilde{p}_\textrm{str}(\boldsymbol{k},\epsilon)=\cfrac{(\epsilon^{\alpha-1}/\Lambda^\alpha)}{(\epsilon/\Lambda)^\alpha
	  + \cfrac{2\Bigl(v_0/2\Lambda\Bigr)^2\boldsymbol{k}^{2}}{(\epsilon/\Lambda)^{\alpha}
          + \cfrac{\Bigl(v_0/2\Lambda\Bigr)^{2}\boldsymbol{k}^{2}}{(\epsilon/\Lambda)^{\alpha}
          + \cfrac{\Bigl(v_0/2\Lambda\Bigr)^{2}\boldsymbol{k}^{2}}{(\epsilon/\Lambda)^{\alpha}+\ddots
          } } }
          },
\end{equation}
this expression is universal in that does not depend on the pattern of active motion. 

If expression~\eqref{p0-ShortTR} is truncated to the first approximant, after some rearrangements we have that $\widetilde{p}_\textrm{str}$ becomes
\begin{equation}\label{pdf fourier lalpace}
\widetilde{p}_\textrm{fWE}(\boldsymbol{k},\epsilon)=
\frac{\epsilon^{2\alpha-1}}{\epsilon^{2\alpha}+\Lambda^{2\alpha}\Bigl(v_0^2/2\Lambda^2\Bigr)\boldsymbol{k}^{2}}
\end{equation}
which corresponds to the solution of the \emph{fractional wave equation} [see Ref.~\cite{schneider1989fractional}], i.e. it satisfies
\begin{equation}\label{fWEq}
\frac{\partial^{2\alpha}}{\partial t^{2\alpha}}p_\textrm{fWE}(\boldsymbol{x},t)=\Lambda^{2\alpha}\frac{v_0^2}{2\Lambda^2}\nabla^2 p_\textrm{fWE}(\boldsymbol{x},t).
\end{equation}
The solution of Eq.~(\ref{fWEq}) can be obtained by inverse Laplace transform of Eq.~(\ref{pdf fourier lalpace}), which yields
\begin{align}
\widetilde{p}_\textrm{fWE}(\boldsymbol{k},t)=E_{2\alpha}\left(-\frac{v_0^2}{2\Lambda^2}\boldsymbol{k}^{2}\Lambda^{2\alpha}t^{2\alpha}\right),
\end{align}
where $E_{\rho}(z)$, is the one-parameter Mittag-Leffler function, see Eq.~(\ref{ML three}) with $\delta=\beta=1$. The inverse Fourier transform of $\widetilde{p}_\textrm{fWE}(\boldsymbol{k},t)$ can be carried out in terms of the Fox $H$-function $H_{p,q}^{m,n}(z)$~(see Eq. (\ref{H_integral}) for a definition). Due to the dependence on $\boldsymbol{k}^2=k^2$, rotational symmetry is manifest and we have 
\begin{subequations}
\begin{equation}\label{fWE-SolutionNew}
p_\textrm{fWE}(\boldsymbol{x},t)
=\int_{0}^{\infty}dk\, kJ_{0}(kx)E_{2\alpha}\left(-\frac{v_0^2}{2\Lambda^2}k^{2}\Lambda^{2\alpha}t^{2\alpha}\right),
\end{equation}
where integration over $\theta$, the angular part of the element $d^2k=kdkd\theta$ has been carried out in the double integral that defines the symmetric inverse Fourier transform. The connection between the Mittag-Leffler function and the Fox $H$-function $H_{p,q}^{m,n} (z)$~Eq. \eqref{HML}, allows for the identification
\begin{multline}\label{M-L-Fox}
 E_{2\alpha}\left(-\frac{v_0^2}{2\Lambda^2}k^{2}\Lambda^{2\alpha}t^{2\alpha}\right)=\\
 H_{1,2}^{1,1}\left[\frac{v_0^2}{2\Lambda^2}k^{2}\Lambda^{2\alpha}t^{2\alpha}\left|\begin{array}{c
l}
     (0,1)\\
    (0,1), (0,2\alpha)
  \end{array}\right.\right],   
\end{multline}
we recognise Eq. \eqref{fWE-SolutionNew} as the Hankel transform of the Fox $H$-function \eqref{hankel transform h} given in \eqref{M-L-Fox}, and after use of the properties \eqref{H_property02} through \eqref{H_property-a}, we find 
\begin{equation}
p_\textrm{fWE}(\boldsymbol{x},t)=\frac{1}{x^2}H_{1,2}^{2,0}\left[\frac{\Lambda}{\sqrt{2}v_0}\frac{x}{(\Lambda t)^\alpha
}\left|\begin{array}{c
l}
     (1,\alpha)\\
    (1,1/2), (1,1/2)
  \end{array}\right.\right].
\end{equation}
\end{subequations}
Thus the coherent motion described by Eq. \eqref{p0-ShortTR} differs from the one given by the fractional wave equation. In such a limit the continuous fraction Eq.~\eqref{p0-ShortTR} can be written as the series expansion
\begin{multline}
\widetilde{p}_\textrm{str}(\boldsymbol{k},\epsilon)=\frac{1}{\epsilon}\biggl\{1-2\biggl(\frac{{v}_0}{2\Lambda}\biggr)^2\frac{\boldsymbol{k}^2}{\epsilon^{2\alpha}/\Lambda^{2\alpha}}+2\biggl[\Bigl(\frac{{v}_0}{2\Lambda}\Bigr)^2\frac{\boldsymbol{k}^2}{\epsilon^{2\alpha}/\Lambda^{2\alpha}}\biggr]^2\\
-2\biggl[\Bigl(\frac{{v}_0}{2\Lambda}\Bigr)^2\frac{\boldsymbol{k}^2}{\epsilon^{2\alpha}/\Lambda^{2\alpha}}\biggr]^3+\ldots\biggr\}
\end{multline}
and after some rearrangements as
\begin{equation}
\widetilde{p}_\textrm{str}(\boldsymbol{k},\epsilon)=\frac{2}{\epsilon}\sum_{n=0}^{\infty}(-1)^n \biggl[\Bigl(\frac{{v}_0}{2\Lambda}\Bigr)^2\frac{\boldsymbol{k}^2}{\epsilon^{2\alpha}/\Lambda^{2\alpha}}\biggr]^n-\frac{1}{\epsilon}.  
\end{equation}
This last expression is susceptible to Laplace inversion, thus we have in Fourier and time variables
\begin{align}
p_\textrm{str}(\boldsymbol{k},t)&=2\sum_{n=0}^\infty\frac{(-1)^n}{\Gamma(2n\alpha+1)}\biggl[\Bigl(\frac{{v}_0}{2}k\Lambda^{\alpha-1}t^\alpha\Bigr)^2\biggr]^n-1\nonumber\\
&=2E_{2\alpha}\biggl[-\Bigl(\frac{{v}_0}{2}k\Lambda^{\alpha-1}t^\alpha\Bigr)^2\biggr]-1,
\end{align}
where $E_{\rho}(z)$ is the one-parameter Mittag-Leffler function, see relation~(\ref{ML three}).

We close this section with some comments. First, the appearance of the fTDE~\eqref{fTDE} and fTTE~\eqref{fTTEq} as particular cases of the reduced description of fractional active motion~\eqref{FRT-Equation}, suggests that the generalised fractional diffusion equation~\eqref{Solution-D-InvLF} belongs to the same class of persistent random walks. Indeed, the description of persistent motion within the \emph{continuous-time random walks} (CTRW) framework of Montroll and Weiss, has allowed the derivation of the standard telegrapher's equation and its fractional generalisation in 2 and dimensions~\cite{masoliver2020fractional2d,masoliver2017fractional3d,masoliver2021review}. In Refs.~\cite{barkai2002ctrwfrac,barkai2003actrw}, the validity of the fractional diffusion equation was explored as the asymptotic limit of the exact solution of the CTRW. The standard two-dimensional telegrapher's equation was obtained in~\cite{masoliver2020fractional2d} as the fluid limit (long times and large distances), of a two-dimensional isotropic and uniform random walk model. Thus studying the exact link between Eq. (\ref{Solution-D}) and the corresponding persistent CTRW approach is of interest and the analysis will be presented elsewhere. 

Our second comment addresses the further generalisation of the transport equation of active motion to include space fractional space derivatives. A straightforward way is the replacement $\vert\boldsymbol{k}\vert^2\rightarrow \vert\boldsymbol{k}\vert^{2\gamma}$, $0<\gamma<1$, in Eq. (\ref{Solution-D}), which modifies the Laplacian in the corresponding generalised diffusion equation (\ref{Solution-D-InvLF}) by introducing the so-called fractional Riesz-Feller derivative operator defined as the Fourier transform of $|\boldsymbol{k}|^{2\gamma}$. A similar generalisation is analysed by Masoliver from the CTRW framework, that stems from the same substitution in the description of anomalous diffusion in \cite{barkai2002ctrwfrac}. This fractional space derivative has been used to model processes such as Lévy flights \cite{barkai2002ctrwfrac}, that when introduced in our case, we would have a model for active particles performing Lévy flights under slowed down dynamics. 

From the generalisation commented in the last paragraph, we can trace back the analysis to find that the transport equation of active motion \eqref{Transformed} is modified accordingly by replacing the term $iv_0\,(\hat{\boldsymbol{v}}\cdot \boldsymbol{k})$ with $iv_{\gamma}|\boldsymbol{k}|^{\gamma}(\hat{\boldsymbol{v}}\cdot \hat{\boldsymbol{k}})$ where $\hat{\boldsymbol{k}}$ points in the direction of the wave vector. Thus, after returning to space representation, the projection of the gradient in the swimmer's direction $v_{0}\hat{\boldsymbol{v}}\cdot \nabla \mathcal{P}$  changes to $v_{\gamma} (\boldsymbol{D}^{\gamma}_{\hat{\boldsymbol{v}}} \mathcal{P})$, where a version of the fractional directional derivative is introduced as $\boldsymbol{D}^{\gamma}_{\hat{\boldsymbol{v}}} \mathcal{P}=\hat{\boldsymbol{v}}\cdot \hat{\boldsymbol{e}}^i \nabla^{\gamma}_i\mathcal{P}$, where $\hat{\boldsymbol{e}}^i$ are orthogonal unit vectors with $i=1,2$, and $\nabla^{\gamma}_i$ is the fractional Riesz-Feller derivative of order $\gamma$ with respect to coordinate $x_i$. Although this prescription for the fractional gradient is similar to the one given by Tarasov in \cite{tarasov2008max}, in there, just the Riemman-Liouville and Caputo derivatives were used. Moreover, the fractional directional derivative for the Riesz-Feller operator was introduced in \cite{meerschaert2006,dovidio2014ad} and differs somehow from the definition needed in this case. Since there are many definitions of these operators in the literature, further study of these terms will be done in a future contribution; for the purposes of this paper, only $\gamma=1$ will be considered.

\section{Effects of the fractional order $\alpha$ on the moments of $p_0(x,t)$}
The time dependence of the moments of $p_0(\boldsymbol{x},t)$ can be determined from the inverse Laplace transform of the moments computed directly from Eq. \eqref{p0}. Let us first consider the second moment or mean squared displacement.

\subsection{Mean squared displacement (MSD)}

The Laplace transform of the mean squared displacement $\langle\widetilde{\boldsymbol{x}^{2}}(\epsilon)\rangle$ is given by
\begin{equation}
\langle\widetilde{\boldsymbol{x}^{2}}(\epsilon)\rangle=-\Bigl[\nabla_{\boldsymbol{k}}^2\, \widetilde{p}_0(\boldsymbol{k},\epsilon)\Bigr]_{\boldsymbol{k}=0}
\end{equation}
due to the rotational symmetry in $\boldsymbol{k}$ space we have simply
\begin{subequations}
\begin{align}\label{msdt}
\langle\widetilde{\boldsymbol{x}^{2}}(\epsilon)\rangle=&-\biggl[\frac{1}{k}\frac{\partial}{\partial k}\biggl(k\frac{\partial}{\partial k}\biggr) \widetilde{p}_0(\boldsymbol{k},\epsilon)\biggr]_{\boldsymbol{k}=0}\\
\label{msd-connecting}
=&\frac{v_{0}^2}{\Lambda^2}\frac{\Lambda^\alpha}{\epsilon^{\alpha+1}}\bigl[\widetilde{\mathfrak{D}}_\alpha(\boldsymbol{k},\epsilon)\bigr]_{\boldsymbol{k}=0}\\
=&\frac{v_{0}^2}{\Lambda^2}\frac{\Lambda^\alpha}{\epsilon^{\alpha+1}}\widetilde{\mathfrak{D}}_\textrm{fTTE}(\epsilon),
\label{MSD-Laplace1}
\end{align}
\begin{figure}
\includegraphics[width=\columnwidth]{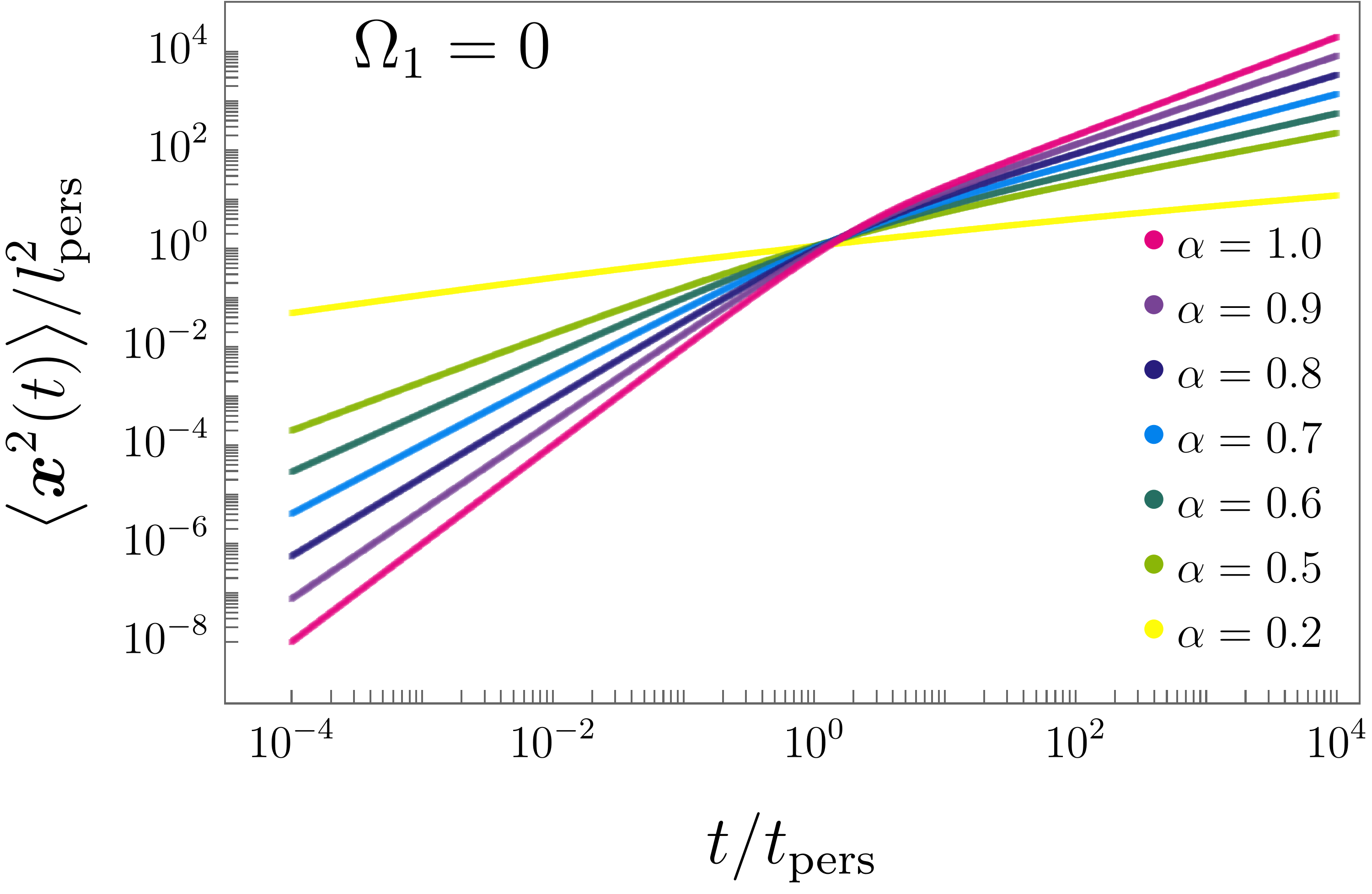}
\includegraphics[width=\columnwidth]{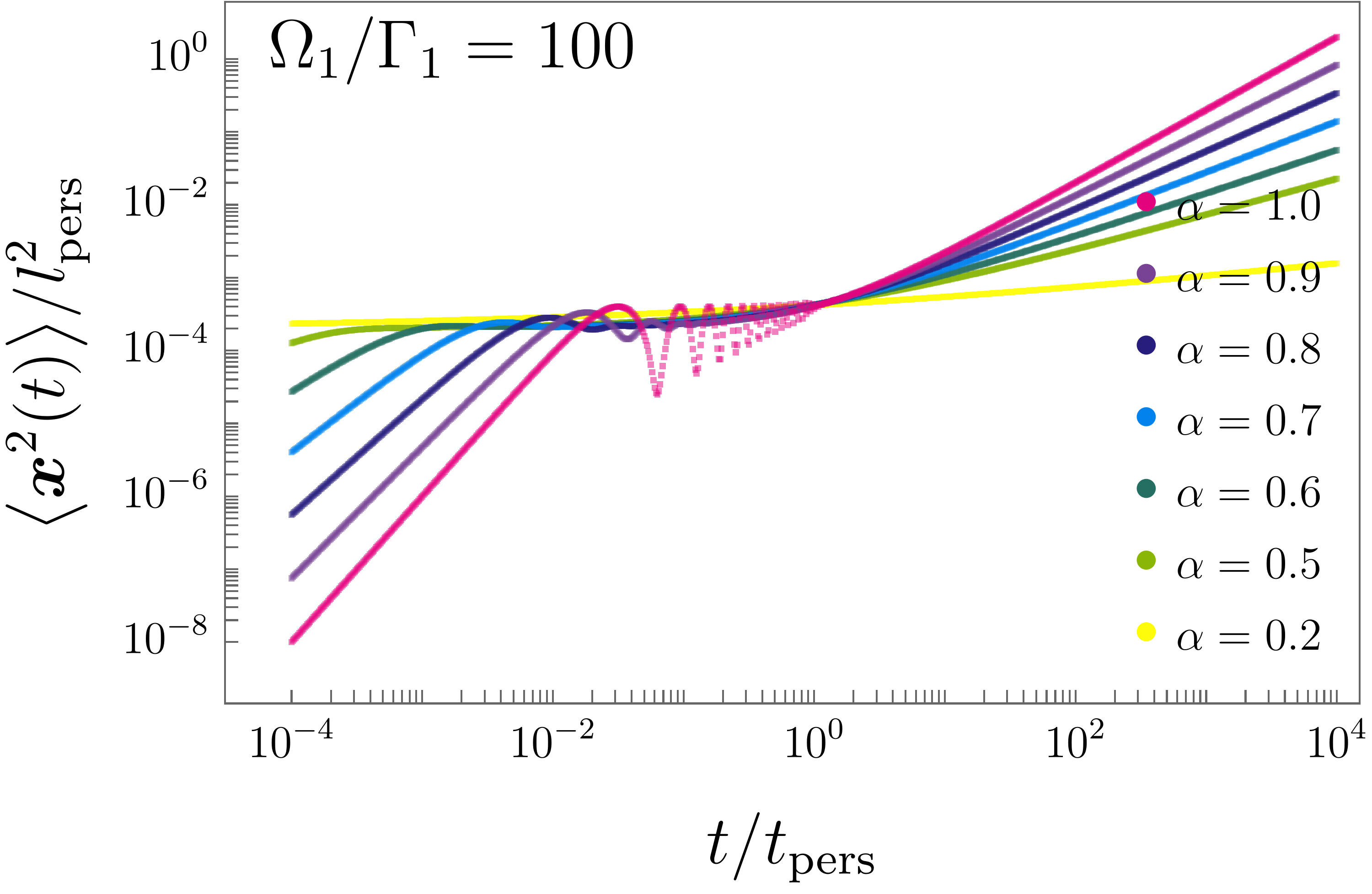}
\caption{Time dependence of the mean squared displacement for different values of order of the fractional derivative $\alpha$. The cases $\alpha= 0.2$, 0.5, 0.6, 0.7, 0.8, 0.9 and 1 are shown and the persistence time $t_\textrm{pers}=(\Lambda\Gamma_1)^{-1}$. Top panel.- The case without rotational bias $\Omega_1=0$. The crossover from the scaling $t^{2\alpha}$ to $t^\alpha$ is apparent. Bottom panel.- The case for which the rotational bias $\Omega_1=100\, \Gamma_1$.  A slowing-down of the time dependence is apparent due to rotational bias ($\Omega_1\neq0$). Oscillations are markedly observed in the case $\alpha=1$, but they are diminished as $\alpha$ decreases, however the slowing down is exacerbated.}
\label{fig:MSD}
\end{figure}
\end{subequations}
where $\widetilde{\mathfrak{D}}_\textrm{fTTE}(\epsilon)$ has appeared already in the previous section and is given in Eq.  \eqref{1stApproximantLaplace}. This connection elucidates that the exact time dependence of the mean squared displacement of persistent motion can be obtained from the first order approximation given in the form of a telegrapher equation.

Explicit Laplace inversion for the expression~\eqref{MSD-Laplace1} is possible in terms of the two-parameters Mittag-Leffler
\begin{multline}\label{msd-time}
\langle\boldsymbol{x}^2(t)\rangle=\frac{v_0^2}{\Lambda^2}\Lambda^{2\alpha}t^{2\alpha}\Bigl[E_{\alpha,2\alpha+1}\bigl(-\lambda_1 \Lambda^\alpha t^{\alpha}\bigr)\\
+E_{\alpha,2\alpha+1}\bigl(-\lambda_1^* \Lambda^\alpha t^{\alpha}\bigr)\Bigr]. 
\end{multline}
For the symmetric scattering functions $Q(\varphi)$ around $\varphi=0$ we get \begin{equation}
\langle\widetilde{\boldsymbol{x}^{2}}(\epsilon)\rangle=2\frac{v_{0}^2}{\Lambda^2}\frac{\Lambda^{2\alpha}}{\epsilon^{1+\alpha}}\frac{1}{\epsilon^\alpha+\Lambda^\alpha\Gamma_1}
\end{equation}
derived from the fractional persistent walk in Ref.~\cite{masoliver2016fractional}. The time dependence is given by
\begin{equation}
\langle\boldsymbol{x}^2(t)\rangle=2\frac{v_{0}^2}{\Lambda^2} \Lambda^{2\alpha}t^{2\alpha}E_{\alpha,2\alpha+1}\bigl(-\Gamma_1\Lambda^\alpha t^{\alpha}\bigr).
\end{equation}
This last expressions exhibits a crossover from the time scaling in the short-time regime
\begin{subequations}\label{MSD-time}
\begin{equation}\label{MSD-STR}
\langle\boldsymbol{x}^2(t)\rangle\approx 2\frac{v_0^{2}}{\Lambda^2}\frac{\Lambda^{2\alpha}}{\Gamma(2\alpha+1)}t^{2\alpha},
\end{equation}
to the long-time regime scaling
\begin{equation}
\langle\boldsymbol{x}^2(t)\rangle\sim 2\frac{v_0^2}{\Lambda^2 }\frac{\Lambda^\alpha }{\Gamma_1\,\Gamma(\alpha+1)}t^\alpha,
\end{equation}
\end{subequations}
as is shown in the top panel of Fig. \ref{fig:MSD} for $\alpha=0.2$, 0.5, 0.6,  0.7, 0.8, 0.9 and 1. 

In the general case when a rotational bias is present, i.e. $\Omega_1\neq0$,
the short-time regime satisfies the same scaling given by \eqref{MSD-STR}, however the scaling at the long-time regime is modified by the appearance of $\Omega_1$. Indeed, as mentioned in the previous section, when $\Omega_1\neq0$, $\mathfrak{D}_\textrm{fTTE}(t)$ can not be factorized into factors that contain either $\Gamma_1$ or $\Omega_1$ only. As consequence the time dependence of the MSD in the long-time regime exhibits the involvement between the persistence and the rotational bias, namely,
\begin{equation}
\langle\boldsymbol{x}^2(t)\rangle\sim 2\frac{v_0^2}{\Lambda^2 }\frac{\Lambda^\alpha \Gamma_1}{(\Gamma_1^2+\Omega_1^2)\,\Gamma(\alpha+1)}t^\alpha.
\end{equation}
Thus the rotational bias slows down the growing with time of the MSD, but without modification of the scaling exponent as is shown in the bottom panel of Fig. \ref{fig:MSD} for $\Omega_1/\Gamma_1=100$. The well-marked oscillations in the case $\alpha=1$ can be noticed \cite{SevillaPRE2020}, these are responsible of the MSD slowing down and can be interpreted as a trapping effect due to rotational bias. For $\Omega_1\neq0$ the oscillations diminish rapidly with $\alpha$, but the self-trapping effects are augmented. 

\subsection{Kurtosis}

The kurtosis of $p_0(\boldsymbol{x}, t)$, $\kappa(t)$, is given by \cite{Mardia74p115}
\begin{equation}\label{kurtosis}
\kappa(t) =\left\langle \left[\bigl(\boldsymbol{x}(t)-\langle \boldsymbol{x}(t)\rangle\bigr)\Sigma^{-1}\bigl(\boldsymbol{x}(t)-\langle 
\boldsymbol{x}(t)\rangle\bigr)^{\text{T}}\right]
^{2}\right\rangle,
\end{equation}
where $\boldsymbol{x}^{\text{T}}$ denotes the transpose of the vector $\boldsymbol{x}$ and $\Sigma $ is the $2\times2$ matrix defined by the average of the dyadic product 
$\bigl(\boldsymbol{x}(t)-\langle \boldsymbol{x}(t)\rangle\bigr)^{\text{T}}\cdot \bigl(\boldsymbol{x}(t) -\langle \boldsymbol{x}(t)\rangle\bigr).$ 
The kurtosis $\kappa(t)$ serves as a measure of the distance of $p_0(\boldsymbol{x},t)$ from the two-dimensional Gaussian distribution, whose kurtosis in two dimensions has the value 8.
Due to the rotational invariance of  $p_0(\boldsymbol{x},t)$ the kurtosis results to be
\begin{equation}\label{kurtsym}
\kappa(t) =4\frac{\langle x^{4}(t)\rangle _{\text{rad}}}{\langle x^{2}(t)\rangle _{\text{rad}}^{2}}, 
\end{equation}
where the even moments of the radial distribution $xp_0(x,t)$ are:   $\langle x^{2n}(t)\rangle_\text{rad}=\int_0^\infty dx\, x^{2n}\, xp_0(x,t)$. The radial distribution is obtained from $p_0(\boldsymbol{x},t)$ through integration over all directions of the position vector when using polar coordinates $\boldsymbol{x}=x(\cos\theta,\sin\theta)$. These moments can be computed in standard manner from the characteristic function given by Eq. \eqref{p0}, thus it is straightforward to compute the Laplace transform of the fourth moment
\begin{align}\label{FourthMomentLaplace}
\langle\widetilde{x^{4}}(\epsilon)\rangle_{\text{rad}}&=\frac{8}{3}\biggl[\frac{\partial^4}{\partial k^4}\widetilde{p}_0
(k,\epsilon)\biggr]_{k=0}.
\end{align}
This in turn, can be written in terms of the connecting function $\widetilde{\mathfrak{D}}_\alpha(\boldsymbol{k},\epsilon)$ as
\begin{subequations}
\begin{multline}\label{FourthMomentLaplace2}
 \langle\widetilde{x^{4}}(\epsilon)\rangle _{\text{rad}}=4\frac{v_0^4}{\Lambda^4}\frac{\Lambda^{2\alpha}}{\epsilon^{2\alpha+1}}\left[{\widetilde{\mathfrak{D}}_\alpha}^2(\boldsymbol{k},\epsilon)\right]_{k=0}\\
 -8\frac{v_0^2}{\Lambda^2}\frac{\Lambda^{\alpha}}{\epsilon^{\alpha+1}}\left[\frac{\partial^{2}}{\partial k^{2}}\widetilde{\mathfrak{D}}_\alpha(\boldsymbol{k},\epsilon)\right]_{k=0},
\end{multline}
we have 
\begin{equation*}
\left[\widetilde{\mathfrak{D}}_\alpha^2(\boldsymbol{k},\epsilon)\right]_{k=0}=\widetilde{\mathfrak{D}}_\textrm{fTTE}^2(\epsilon)
=\biggl(\frac{\Lambda^\alpha}{\epsilon^\alpha+\Lambda^\alpha \lambda_1}
+\frac{\Lambda^\alpha}{\epsilon^\alpha+\Lambda^\alpha\lambda_1^{*}}\biggr)^{2},
\end{equation*}
and
\begin{multline*}
\biggl[\frac{\partial^{2}}{\partial 
k^{2}}\widetilde{\mathfrak{D}}_\alpha(\boldsymbol{k},t)\biggr]_{k=0}=-\frac{v_0^2}{2\Lambda^2}\biggl[\frac{\Lambda^{3\alpha}}{(\epsilon^\alpha+\Lambda^\alpha\lambda_1)^{2}(\epsilon^\alpha+\Lambda^\alpha\lambda_2)}\\+\frac{\Lambda^{3\alpha}}{(\epsilon^\alpha+\Lambda^\alpha\lambda_1^*)^{2}(\epsilon^\alpha+\Lambda^\alpha\lambda_2^*)}\biggr].
\end{multline*}
Thus the fourth moment in Laplace domain is explicitly given by
\begin{multline}\label{4thMoment-L}
\bigl\langle 
\widetilde{x^{4}}(\epsilon)\bigr\rangle_\textrm{rad}=4\frac{v_0^4}{\Lambda^4}\frac{\Lambda^{4\alpha}}{\epsilon^{\alpha+1}}\biggl[\frac{1}{\epsilon^\alpha}\left(\frac{1}{\epsilon^\alpha+\Lambda^\alpha\lambda_1}+\frac{1}{\epsilon^\alpha+\Lambda^\alpha\lambda_1^*}\right)^{2}\\
+\frac{1}{(\epsilon^\alpha+\Lambda^\alpha\lambda_{1})^{2}(\epsilon^\alpha+\Lambda^\alpha\lambda_{2})}\\
+\frac{1}{(\epsilon^\alpha+\Lambda^\alpha\lambda_{1}^{*})^{2}(\epsilon^\alpha+\Lambda^\alpha\lambda_{2}^{*})}\biggr],
\end{multline}
\end{subequations}
which incorporates $\lambda_2$ and $\lambda_2^*$ and therefore incorporates the second order trigonometric moments $\Gamma_2=1-\langle\cos2\varphi\rangle_Q$ and $\Omega_2=\langle\sin2\varphi\rangle_Q$.

Inversion of the Laplace transform of Eqs.~\eqref{4thMoment-L} is possible in terms of the function $\mathcal{E}_{\alpha,\beta}^\delta(t;\eta)=t^{\beta-1}E_{\alpha,\beta}^\delta(-\eta t^\alpha)$ \cite{SandevBook2019}, where  $E_{\rho,\beta}^{\delta}(z)$ is the Prabhakar or three-parameters Mittag-Leffler function, [see Eq.~\eqref{ML three}], namely 
\begin{multline}\label{4thm}
\langle x^{4}(t)\rangle=4\frac{v_0^4}{\Lambda^4}\biggl[\mathcal{E}_{\alpha,4\alpha}^2\bigl(\Lambda t;\lambda_1\bigr)+\mathcal{E}_{\alpha,4\alpha}^2\bigl(\Lambda t;\lambda_1^*\bigr)\\
+2\int_{0}^{t}dt^\prime\mathcal{E}_{\alpha,2\alpha}\bigl[\Lambda(t-t');\lambda_1\bigr]\mathcal{E}_{\alpha,2\alpha}\bigl(\Lambda t^\prime;\lambda_{1}^{*}\bigr)\\
+\int_{0}^{t}dt' \mathcal{E}_{\alpha,4\alpha}^2\bigl[(\Lambda(t-t');\lambda_1\bigr]E_\alpha\bigl(-\lambda_2\Lambda^\alpha {t'}^\alpha\bigr)\\
+\int_{0}^{t}dt'\mathcal{E}_{\alpha,4\alpha}^2\bigl(\Lambda(t-t');\lambda_1^*\bigr)E_{\alpha}\bigl(-\lambda_2^*\Lambda^\alpha{t'}^\alpha\bigr)\biggr].
\end{multline}
Correspondingly, $\langle\widetilde{x^2}(\epsilon)\rangle_\text{rad}$ coincides exactly with expression~\eqref{msd-connecting}, whose inverse Laplace transform is given in Eq.~\eqref{msd-time}. 

From these results, it can be shown that the specific pattern of active motion (carried in $Q(\varphi)$) has no influence on the short- and long-time behaviour of the kurtosis $\kappa(t)$, depending only on $\alpha$ as shown as follows. In the short-time regime we have
\begin{equation}
\bigl\langle x^4(t)\bigr\rangle\approx24\frac{v_0^4}{\Lambda^4}\frac{(\Lambda t)^{4\alpha}}{\Gamma(4\alpha+1)}
\end{equation}
while 
\begin{equation}
\bigl\langle x^2(t)\bigr\rangle\approx2\frac{v_0^2}{\Lambda^2}\frac{(\Lambda t)^{2\alpha}}{\Gamma(2\alpha+1)}    
\end{equation}
leading to 
\begin{equation}
\kappa_\textrm{s-t}\approx24\frac{\bigl[\Gamma(2\alpha+1)\bigr]^2}{\Gamma(4\alpha+1)}.    
\end{equation}

In the long-time regime, from the asymptotic relation~(\ref{GML large z}), we have 
\begin{equation}
\bigl\langle x^4(t)\bigr\rangle\sim4\frac{v_0^4}{\Lambda^4}\biggl(\frac{1}{\lambda_1}+\frac{1}{\lambda_1^*}\biggr)^2 \frac{(\Lambda t)^{2\alpha}}{\Gamma(2\alpha+1)}
\end{equation}
and
\begin{equation}
\bigl\langle x^2(t)\bigr\rangle\approx\frac{v_0^2}{\Lambda^2}\biggl(\frac{1}{\lambda_1}+\frac{1}{\lambda_1^*}\biggr) \frac{(\Lambda t)^{\alpha}}{\Gamma(\alpha+1)}    
\end{equation}
which gives the kurtosis in the long-time regime as
\begin{equation}
\kappa_\textrm{l-t}\sim16\frac{\bigl[\Gamma(\alpha+1)\bigr]^2}{\Gamma(2\alpha+1)}.    
\end{equation}

\begin{figure}
\includegraphics[width=\columnwidth,trim=10 10 0 10, clip=true]{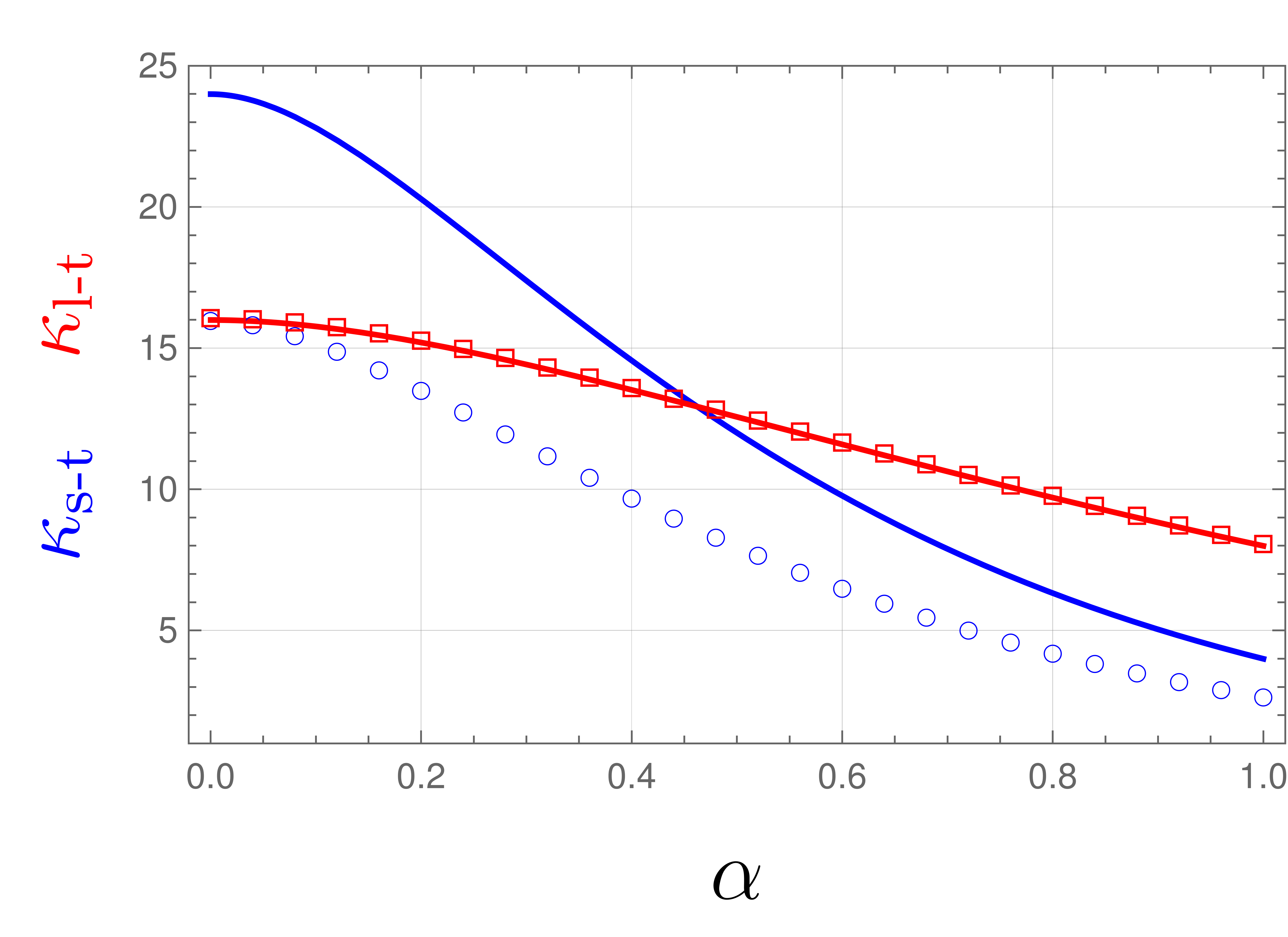}
\caption{The kurtosis in the short- ($\kappa_\textrm{s-t}$ solid-blue line) and long-time ($\kappa_\textrm{l-t}$ solid-red line) regime as function of the order of the fractional derivative $\alpha$. Symbols indicate the corresponding limits obtained from the fractional time telegrapher's equation. The red squares correspond to the long time regime of the fractional time telegrapher's equation and exactly match the red line. The blue circles correspond to the long time limit of the fractional time telegrapher's equation, i.e., the time fractional wave equation.}
\label{fig:kurtosis-Short-LongTime}
\end{figure}

The short and long-time behaviour of the kurtosis as a function of the anomalous exponent is shown in Fig. \ref{fig:kurtosis-Short-LongTime} (solid blue and red lines, respectively). The intersect of both curves at the value $\alpha\approx0.462151$ obtained from 
\begin{equation}
\frac{[\Gamma(\alpha)]^2\Gamma(4\alpha)}{[\Gamma(2\alpha)]^{3}}=3. 
\end{equation}

The long-time behaviour of the kurtosis of fractional active motion coincides with the long-time behaviour of the kurtosis given by the time fractional telegrapher's equation (red line), which is expected since, although the second and fourth moments depend on the specificity of the pattern of active motion, this regime is irrelevant for the kurtosis.

The striking difference in the short-time regime is even more remarkable. Although functionally, the behaviour of both expressions is the same, the coefficient of the one calculated with the solution of the fractional wave equation is $2/3$ of the coefficient of the expression in the same regime but accounting for all approximants of the connecting function. This fact suggests that, even in the regime where the activity pattern model is unimportant, the approximants of the connecting function of the fractional active motion influence the kurtosis' behaviour, which is evident for values of $\alpha$ close to zero.

\subsection{Uniform angle-scattering distribution}

Let us consider again the uniform scattering angle distribution on $[-\pi, \pi]$
\begin{equation}
Q(\varphi)=\frac{1}{2\pi},    
\end{equation}
which is used to model the so-called {\it run-and-tumble} two-dimensional pattern of active motion. This distribution leads to a single time scale, $\Lambda^{-1}$, i.e., $\Gamma_n=1, \forall\; n\ge1$. 
\begin{figure}
\includegraphics[width=\columnwidth,trim=7 5 0 3, clip=true]{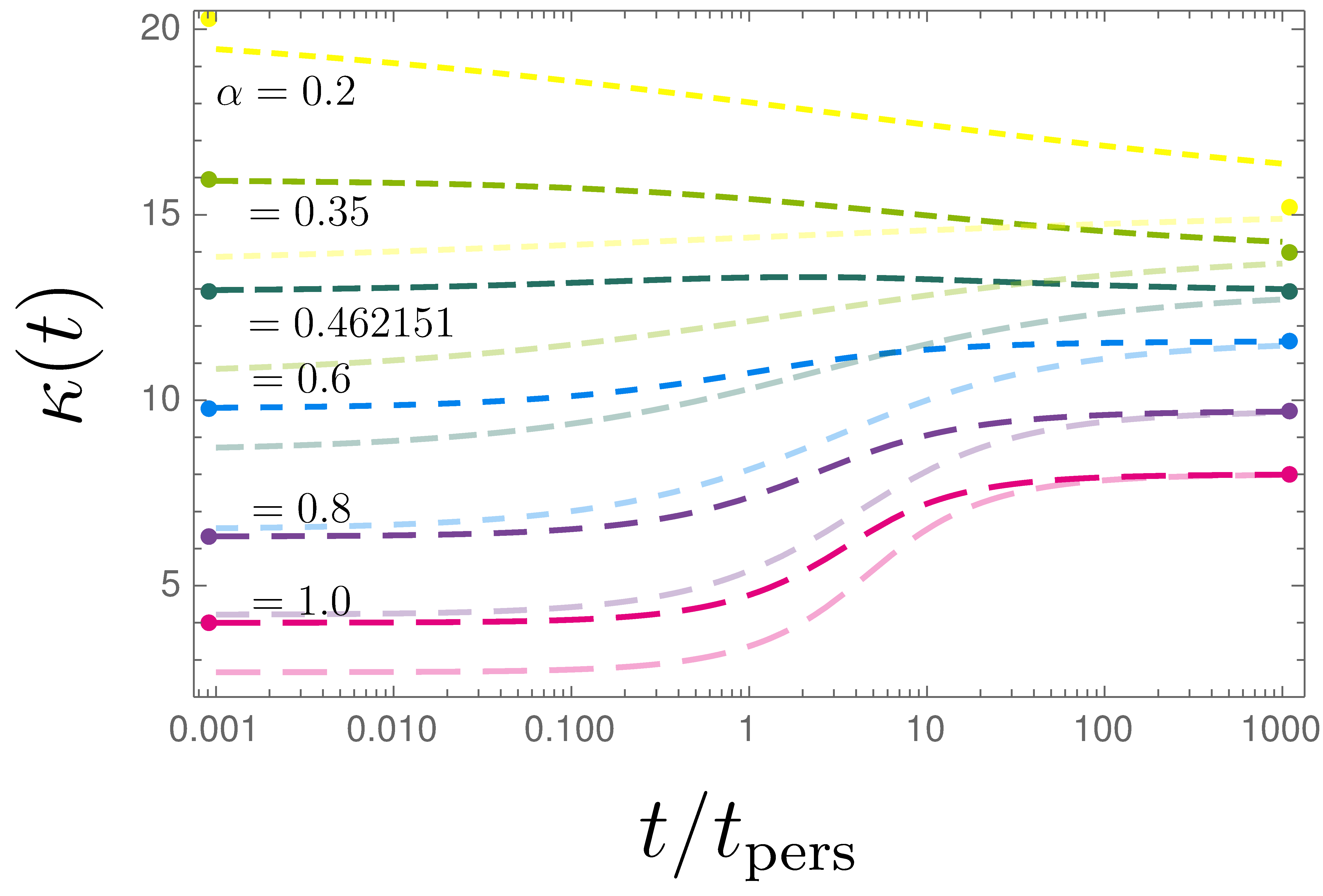}
\caption{Time dependence of the kurtosis, $\kappa(t)$, for the uniform run-and-tumble case, $Q(\varphi)=(2\pi)^{-1}$, for the values $\alpha=0.2$, 0.35, 0.462151, 0.6, 0.8 and 1.0, correspondingly marked from short to long dashes. The dots mark the values in the limit $t\rightarrow\infty$ (right), and those in the limit $t\rightarrow0$ (left).}
\label{fig:Kurtosis-Run-and-Tumble}
\end{figure}

The MSD can be derived from the equation (\ref{msdt}), which reduces to Masoliver's expression \cite{masoliver2016fractional}, for the fractional telegrapher's equation,
\begin{equation}
    \langle\boldsymbol{x}^2(t)\rangle = 2{v}_0^2\,t^{2\alpha}\Lambda^{2(\alpha-1)} E_{\alpha,2\alpha+1}(-{\Lambda^{\alpha}} t^{\alpha}).
\end{equation}
The short-time limit yields $\langle\boldsymbol{x}^2(t)\rangle \sim 2\,\frac{{v}_0^2(\Lambda t)^{2\alpha}}{\Lambda^2\Gamma(2\alpha+1)}$, while the long-time behaviour is of form $\langle\boldsymbol{x}^2(t)\rangle \sim \frac{2{v}_0^2}{{\Lambda}^2}\,\frac{(\Lambda t)^{\alpha}}{\Gamma(\alpha+1)}$.
The fourth moment can also be obtained from~(\ref{4thMoment-L}) when the parameters are the same, i.e. ${\lambda}_1={\lambda}_1^*={\lambda}_2={\lambda}_2^*=1$, yielding the following 
\begin{equation}
   \langle \widetilde{x^{4}}(\epsilon)\rangle = 4{v}_{0}^{4} \Lambda^{4(\alpha-1)}\left[ \frac{4 \epsilon^{-2\alpha-1}}{(\epsilon^\alpha+{\Lambda^{\alpha}})^2} + \frac{2 \epsilon^{-\alpha-1}}{(\epsilon^\alpha+{\Lambda^{\alpha}})^3}\right],
\end{equation}
whose inverse Laplace transform can be written in terms of the $\mathcal{E}_{\alpha,\beta}^{\gamma}$ function 
\begin{align}\label{x4}
   \langle x^{4}(t)\rangle=8\frac{{v}_{0}^{4}}{\Lambda^{4}}\left[ 2 \mathcal{E}_{\alpha,4\alpha+1}^2\bigl(\Lambda t;1\bigr) +  \mathcal{E}_{\alpha,4\alpha+1}^3\bigl(\Lambda t;1\bigr) \right].
\end{align}
Short-time limit yields $\langle x^{4}(t)\rangle\sim \frac{24{v}_{0}^{4}t^{4\alpha}\Lambda^{4(\alpha-1)}}{\Gamma(4\alpha+1)}$, while the long-time limit $\langle x^{4}(t)\rangle\sim \frac{16{v}_{0}^{4}t^{2\alpha}\Lambda^{2(\alpha-2)}}{\Gamma(2\alpha+1)}$. We can compare Eq. (\ref{x4}) with the fourth moment calculated with (\ref{kurtsym}) and the characteristic function of Ref. \cite{masoliver2020fractional2d}, namely,
\begin{align}
    &\langle \widetilde{x^{4}}(\epsilon)\rangle =
    \frac{16 {v}_0^4 \Lambda^{4(\alpha-1)}}{\epsilon^{2 \alpha +1} \left(\epsilon^{\alpha} +{\Lambda^{\alpha}} \right)^2}, \nonumber\\& \langle x^{4}(t)\rangle = 16 {v}_0^4 \Lambda^{-4} \mathcal{E}_{\alpha,4\alpha+1}^2\bigl(\Lambda t;1\bigr).\label{x4tele}
\end{align}
Again, we can see that in the long-time limit, both expressions match, whereas in the short-time regime, the coefficient of the latter case is 1.5 times smaller, indicating that Eq. (\ref{x4}) contains additional information of the connecting function of the active motion ignored by (\ref{x4tele}). The temporal behaviour of the kurtosis is shown in Fig. \ref{fig:Kurtosis-Run-and-Tumble}. We clearly observe that a diffusive behaviour given by the time fractional telegrapher's equation is recovered at long times. In contrast, at short times, there is a difference between fractional telegrapher's equation results and the present fractional active motion approach, mainly due to the structure of the connecting function and its role as a memory kernel in the generalised nonlocal diffusion equation. We also observe that such a difference is accentuated for small values of the anomalous exponent.

\section{Summary and Final Remarks}
We have analysed the fractional-time generalisation of the two-dimensional transport equation for active motion \cite{SevillaPRE2020}. We achieved the generalisation by replacing the time derivative with the Caputo's prescription of the fractional derivative and provided exact results for an arbitrary active motion pattern specified by the scattering function $Q(\varphi)$ of the self-propulsion direction. 

After consideration of the symmetry of the initial pulse, we solved exactly the governing equation in Fourier-Laplace space, such that the characteristic function is expressed in terms of continued fractions Eq.~\eqref{p0} or in a more compact and familiar form as in Eq.~\eqref{Solution-D}, where we introduce a connecting function $\widetilde{\mathfrak{D}}_\alpha(\boldsymbol{k},\epsilon)$. By inverting the Fourier-Laplace transform, we obtained a generalised nonlocal diffusion equation Eq.~\eqref{Solution-D-InvLF}. In such expression, the connecting function plays the role of a memory kernel responsible for the spatial and temporal smearing of the Laplacian operator, and whose nature depends on the scattering-angle distribution ${Q}(\varphi)$. Remarkably, the connecting function explicitly depends on the active swimmer's persistence time and effective period of rotation, together with the multiple time scales $\Gamma_n^{-1}$ and $\Omega_n^{-1}$, for $n\neq1$, that define the statistical features of the motion at all temporal regimes. On the left-hand side, the Caputo fractional derivative appears; it may also come from a power-law memory kernel, modelling the memory of the medium. Our generalisation contains as many refinements as approximants of the connecting function we consider. We studied the extreme short-time and long-time regimes. In the former, it is possible to obtain an expression independent of the angular scattering distribution whose Fourier-Laplace inverse is written in terms of a Mittag-Leffler function. Suppose only the first approximant of the connecting function is used. In that case, the expression reduces to the two-dimensional fractional wave equation whose solution is given by the Fox $H$-function. In the extreme case where the frequency is much smaller than $\Lambda$, i.e., the long-time regime, the connecting function becomes two, and the nonlocal generalised diffusion equation turns into the time-fractional diffusion equation. For the first approximant, the connecting function can be written in terms of Mittag-Leffler functions; when $\alpha=1$, we obtain the previous result from which we can identify the persistence time in the direction of motion and the angular frequency that causes the active swimmer to change direction. In the case of a uniform angular distribution, no angular velocity is present; then, the connecting function collapses to a single factor, yielding the 2D fractional telegrapher's equation (\ref{fTTEq}) upon Laplace-Fourier inversion.

We provided a comprehensive analysis of the effects of active motion and of the fractional exponent on the moments of the distribution of the particle positions; specifically, we calculate expressions for the mean square displacement and kurtosis. The mean square displacement depends only on the terms of the first approximant of the connecting function and written in terms of two-parameter Mittag-Leffler functions. For the case of uniform $Q(\varphi)$, it reduces to the expression previously found from the persistent fractional walk. The MSD shows a time-scale transition in power law from the short- to the long-time regime. The behaviour of the MSD for $Q(\varphi)$ with angular velocity such that $\Omega_1=100\Gamma_1$ exhibits a trapping effect as oscillations with increasing magnitude as the anomalous exponent approaches one. We obtain an expression for the fourth moment in terms of Prabhakar's function. It is interesting how, at extreme time limits, the kurtosis does not depend on the active motion model given by $Q(\varphi)$. In the long-time case, the dependence of the kurtosis on the fractional exponent is the same as for the fractional telegrapher's fractional equation. Conversely, at short times, there is a considerable difference, quite notorious for small alpha, between the behaviour of the kurtosis from the fractional telegrapher's equation and the kurtosis of the fractional active motion, in the latter arising from the connection function in its role as a memory kernel.

We have approached the nontrivial effects of the fractional dynamics to the nonequilibrium dynamics of active motion, this allowed us to provide a description of the anomalous diffusion of self-propelled particles (that self-propelled with an arbitrary pattern motion) observed in different studies. Further analysis contemplates deepening on the study of the Riezs-Feller gradient for the fractional active transport equation and the analysis of specific scattering functions that takes into account the variety of the in biological organisms \cite{TaktikosPlos2014}. Current research is taking into account more and distinctive nonequilibrium effects to active motion, as \emph{stochastic resetting} \cite{EvansPRL2011}. This studies will provide either an advancement to our comprehension of nonequilibrium active motion or the foresight of potential applications of the resulting effects, these will be presented elsewhere.  

\begin{acknowledgments}
This work was supported by UNAM-PAPIIT IN112623. GCA acknowledges partial support form {\it Becas UAM de  Superaci\'on Acad\'emica Elisa Acuña}. TS acknowledges financial support by the German Science Foundation (DFG, Grant number ME~1535/12-1). TS is supported by the
Alliance of International Science Organizations (Project No.~ANSO-CR-PP-2022-05). TS is also supported by the Alexander von Humboldt Foundation. 
\end{acknowledgments}

\appendix

\section{The Mittag-Leffler function and the Fox $H$-function}\label{appA}

The three-parameter Mittag-Leffler function is defined by~\cite{prabhakar1971singular}
\begin{align}\label{ML three}
    E_{\rho,\beta}^{\delta}(z)=\sum_{n=0}^{\infty}\frac{(\delta)_{n}}{\Gamma(\rho n+\beta)}\frac{z^{n}}{n!}.
\end{align} 
where $(\delta)_n=\Gamma(\delta+n)/\Gamma(\delta)$ is the Pochhammer symbol. Its Laplace transform reads
\begin{align}\label{laplaceML}    \mathcal{L}\left[t^{\beta-1}E_{\rho,\beta}^{\delta}\left(-\lambda{t}^{\rho}\right)\right]=\frac{\epsilon^{\rho\delta-\beta}}{(\epsilon^\rho+\lambda)^\delta},
\end{align}
where $|\lambda/\epsilon^{\rho}|<1$. Note that for $\delta=1$ it becomes the two-parameter Mittag-Leffler function, $E_{\rho,\beta}^{1}(z)=E_{\rho,\beta}(z)$, and for $\beta=\delta=1$ it becomes the one-parameter Mittag-Leffler function, $E_{\rho,1}^{1}(z)=E_{\rho,1}(z)=E_{\rho}(z)$. Here we note that since in the three-parameter Mittag-Leffler function parameter $\delta$ is a positive integer, the results can be represented as derivatives of two-parameter Mittag-Leffler function due to the relation~\cite{kilbas2006theory}
\begin{align}\label{two vs three ml}
    \left(\frac{d}{dz}\right)^{n}E_{\alpha,\beta}(z)=n!\,E_{\alpha,\beta+\alpha n}^{n+1}(z), \quad n\in \mathbb{N}.
\end{align}
Moreover, if lambda parameters under the integral are equal the result can be simplified due to the following formula~\cite{kilbas2004generalized}
\begin{align}\label{integral ml functions}
        &\int_{0}^{t}t'^{\beta-1}(t-t')^{\mu-1}E_{\alpha,\beta}^{\delta}\left(-\omega\,t'^{\alpha}\right)E_{\alpha,\mu}^{\gamma}\left(-\omega\,(t-t')^{\alpha}\right)dt'\nonumber\\&=t^{\beta+\mu-1}E_{\alpha,\beta+\mu}^{\delta+\gamma}\left(-\omega\,t^{\alpha}\right),
\end{align}
where $\alpha,\beta,\mu,\gamma,\delta\in\mathbb{C}$, $\Re\{\alpha\}>0$, $\Re\{\beta\}>0$, $\Re\{\mu\}>0$.

For $0<\alpha<2$, the following formula is valid~\cite{gorenflo2014mittag,garra2018prabhakar,sandev2023special}
\begin{equation}\label{GML_formula3}
E_{\alpha,\beta}^{\delta}(-z)=\frac{z^{-\delta}}{\Gamma(\delta)}\sum_{n=0}^{\infty}\frac{\Gamma(\delta+n)}{\Gamma(\beta-\alpha(\delta+n))}\frac{(-z)^{-n}}{n!}, \quad z>1.
\end{equation}
Therefore, one finds the asymptotic behavior of the three-parameter Mittag-Leffler function for large $z$ ($z\gg1$)
\begin{equation}\label{GML large z}
E_{\alpha,\beta}^{\delta}(-z)\sim\frac{z^{-\delta}}{\Gamma(\beta-\alpha\delta)}-\gamma\frac{z^{-(\delta+1)}}{\Gamma(\beta-\alpha(\delta+1))}, \quad z\gg1.
\end{equation}
For $z\ll1$, the three-parameter Mittag-Leffler function behaves as
\begin{align}\label{GML small z alpha}
E_{\alpha,\beta}^{\delta}(\pm\lambda t^{\alpha})&\sim\frac{1}{\Gamma(\beta)}\pm\delta\frac{\lambda t^{\alpha}}{\Gamma(\alpha+\beta)},
\quad t\ll1.
\end{align}

The Fox $H$-function is defined by the
following Mellin-Barnes integral~\cite{saxena_book}
\begin{align}\label{H_integral}
H_{p,q}^{m,n}\left[z\left|\begin{array}{c l}
    (a_1,A_1),...,(a_p,A_p)\\
    (b_1,B_1),...,(b_q,B_q)
  \end{array}\right.\right]&=H_{p,q}^{m,n}\left[z\left|\begin{array}{c l}
    (a_p,A_p)\\
    (b_q,B_q)
  \end{array}\right.\right]\nonumber\\&=\frac{1}{2\pi\imath}\int_{\Omega}ds\,\theta(s)z^{s},
\end{align}
where
$$\theta(s)=\frac{\prod_{j=1}^{m}\Gamma(b_j-B_js)\prod_{j=1}^{n}\Gamma(1-a_j+A_js)}{\prod_{j=m+1}^{q}\Gamma(1-b_j+B_js)\prod_{j=n+1}^{p}\Gamma(a_j-A_js)},$$
$0\leq n\leq p$, $1\leq m\leq q$, $a_i,b_j \in C$, $A_i,B_j \in
R^{+}$, $i=1,...,p$, $j=1,...,q$. The contour $\Omega$ starting at
$c-\imath\infty$ and ending at $c+\imath\infty$ separates the poles
of the function $\Gamma(b_j+B_js)$, $j=1,...,m$ from those of the
function $\Gamma(1-a_i-A_is)$, $i=1,...,n$. 

Three-parameter Mittag-Leffler function~(\ref{ML three}) is a special case of Fox $H$-function~\cite{saxena_book}
\begin{align}\label{HML}
E_{\alpha,\beta}^{\delta}(-z)=\frac{1}{\Gamma(\delta)}H_{1,2}^{1,1}\left[z\left|\begin{array}{l}
    (1-\delta,1)\\
    (0,1),(1-\beta,\alpha)
  \end{array}\right.\right], \quad \Re(\delta)>0.
\end{align}

The Mellin transform of the $H$-function is given by~\cite{saxena_book}
\begin{align}\label{integral of H}
&\int_{0}^{\infty}dx\,x^{\xi-1}H_{p,q}^{m,n}\left[ax\left|\begin{array}{c
l}
    (a_p,A_p)\\
    (b_q,B_q)
  \end{array}\right.\right]=a^{-\xi}\,\theta(-\xi),
\end{align}
where $\theta(-\xi)$ is defined in relation (\ref{H_integral}).

The Hankel transform of the Fox $H$-function is given by~\cite{saxena_book}
\begin{align}\label{hankel transform h}
&\int_{0}^{\infty}dx\,x^{\rho-1}J_{\nu}(ax)H_{p,q}^{m,n}\left[bx^{\sigma}\left|\begin{array}{c
l}
     (a_p,A_p)\\
    (b_q,B_q)
  \end{array}\right.\right]\nonumber\\
&=\frac{2^{\rho-1}}{a^\rho}\nonumber\\&\times H_{p+2,q}^{m,n+1}\left[b\left(\frac{2}{a}\right)^{\sigma}\left|\begin{array}{c l}
     (1-\frac{\rho+\nu}{2},\frac{\sigma}{2}), (a_p,A_p), (1-\frac{\rho-\nu}{2},\frac{\sigma}{2})\\
    (b_q,B_q)
  \end{array}\right.\right].
\end{align}

The following reduction formulas hold true, for $n\geq1$, $q>m$,
\begin{align}\label{H_property0}
&H_{p,q}^{m,n}\left[z\left|\begin{array}{c l}
    (a_1,A_1),...,(a_p,A_p)\\
    (b_1,B_1),...,(b_{q-1},B_{q-1}),(a_1,A_1)
  \end{array}\right.\right]\nonumber\\&=H_{p-1,q-1}^{m,n-1}\left[z\left|\begin{array}{c l}
    (a_2,A_2),...,(a_p,A_p)\\
    (b_1,B_1),...,(b_{q-1},B_{q-1})
  \end{array}\right.\right],
\end{align}
and
\begin{align}\label{H_property02}
&H_{p,q}^{m,n}\left[z\left|\begin{array}{c l}
    (a_1,A_1),...,(a_{p-1},A_{p-1}),(b_1,B_1)\\
    (b_1,B_1),...,(b_q,B_q)
  \end{array}\right.\right]\nonumber\\&=H_{p-1,q-1}^{m-1,n}\left[z\left|\begin{array}{c l}
    (a_1,A_1),...,(a_{p-1},A_{p-1})\\
    (b_2,B_2),...,(b_{q},B_{q})
  \end{array}\right.\right].
\end{align}

The Fox $H$-function has the following properties, as well: 
\begin{align}\label{H_property3}
H_{p,q}^{m,n}\left[z\left|\begin{array}{c l}
    (a_p,A_p)\\
    (b_q,B_q) \end{array}\right.\right]=H_{q,p}^{n,m}\left[\frac{1}{z}\left|\begin{array}{c l}
    (1-b_q,B_q)\\
    (1-a_p,A_p)
  \end{array}\right.\right].
\end{align}
For $\delta>0$, it holds 
\begin{align}\label{H_property-a}
H_{p,q}^{m,n}\left[z^{\delta}\left|\begin{array}{l}
    (a_p,A_p)\\
    (b_q,B_q)
  \end{array}\right.\right]=\frac{1}{\delta} H_{p,q}^{m,n}\left[z\left|\begin{array}{c l}
    (a_p,A_p/\delta)\\
    (b_q,B_q/\delta)
  \end{array}\right.\right]
\end{align}
and 
\begin{align}\label{H_property2}
z^{\sigma}H_{p,q}^{m,n}\left[z\left|\begin{array}{c l}
    (a_p,A_p)\\
    (b_q,B_q)
  \end{array}\right.\right]=H_{p,q}^{m,n}\left[z\left|\begin{array}{c l}
    (a_p+\sigma A_p,A_p)\\
    (b_q+\sigma B_q,B_q)
  \end{array}\right.\right].
\end{align}

\end{document}